# Direct Determination of the Topological Thermal Conductance via Local Power Measurement

Ron Aharon Melcer[1], Sofia Konyzheva[1,2], Moty Heiblum[1], and Vladimir Umansky[1].

1. *Braun Center for Submicron Research, Department of Condensed Matter Physics, Weizmann Institute of Science, Rehovot 761001, Israel*
2. *Laboratory of Quantum Materials (QMAT), Institute of Materials (IMX), École polytechnique fédérale de Lausanne, 1015 Lausanne, Switzerland*

**Thermal conductance measurements, sensitive to charge and chargeless energy flow, are evolving as an essential measurement technique in Condensed Matter Physics. For two-dimensional topological insulators, the measurements of the thermal Hall conductance, $\kappa_{xy}T$, and the longitudinal one $\kappa_{xx}T$, are crucial for the understanding of their underlying topological order. Such measurements are thus far lacking, even in the extensively studied quantum Hall effect (QHE) regime. Here, we report a new local power measurement technique that reveals the topological thermal Hall conductance (not the ubiquitous two-terminal one). For example, we find $\kappa_{xy} \sim 0$ of the challenging $\nu = \frac{2}{3}$ particle-hole conjugated state. This is in contrast to the two-terminal measurement, which provides a non-universal value that depends on the extent of thermal equilibration between the counter-propagating edge modes. Moreover, we use this technique to study the power carried by the current fluctuations in a partitioned edge mode with an out-of-equilibrium distribution.**

The importance of heat flow in electronic systems is being appreciated in recent years, as a growing number of thermal transport[1, 2] and microscopy[3, 4] techniques are being developed. The significance of thermal transport is especially apparent in topologically non-trivial materials. For two-dimensional topological insulators, the bulk of the electron gas is insulating, while gapless excitations flow in 1D-like chiral or helical modes next to the sample's edge[5, 6, 7]. The nature of the edge modes is a manifestation of the wavefunction in the bulk (due to 'bulk-boundary' correspondence[8]), making edge transport experiments a compelling route for studying bulk properties[9]. In addition to the quantization of the electrical conductance of the edge modes, the thermal Hall conductance $\kappa_{xy}T$ ($T$ the temperature) is also expected to be quantized in units of the thermal conductance quanta $\kappa_0 T = \frac{\pi^2 k_b^2}{3h} T$ ($k_b$ is the Boltzmann constant and $h$ is the Planck's constant)[10]. Sensitive to all types of edge modes, the thermal measurement supplements the electrical conductance, and can distinguish between competing topological phases[11]. The importance of the thermal Hall conductance is particularly apparent in topological superconductors, where a half quantized $\kappa_{xy}$ is a signature of the chiral Majorana mode[12] (which does not contribute to the electrical conductance).

The most diverse 'play-ground' for topological condensed matter physical phenomena is the quantum Hall Effect (QHE). The signature of QHE states is their quantized electrical Hall conductance: $G_{xy} = \nu G_0$, where $G_0 = \frac{e^2}{h}$ is the conductance quantum ($e$ the electron's charge), and $\nu$ is the Landau level filling, which can be an integer or a simple fraction. Similarly, their thermal Hall conductance coefficient is quantized: $\kappa_{xy} = \nu_Q \kappa_0$, with $\nu_Q$ an integer for abelian QHE states and a fraction for non-abelian QHE states[11]. For abelian states, the value of $\nu_Q$ is given by the net number of chiral edge modes, $\nu_Q = n_d - n_u$, where $n_d$ ($n_u$) is the number of downstream (upstream) propagating modes (downstream being the chirality dictated by the magnetic field)[10, 13]. The value of $\nu_Q$ (in analogy to $\nu$) is an inherent property of the bulk's topological order[8, 13], and is independent of the exact details of the edge structure.

Measuring $\kappa_{xy}$ is essential when studying a QHE state whose topological order is not known. Its value can distinguish between abelian and non-abelian topological orders, and even discriminate between competing non-abelian orders[11]. So far, the thermal conductance of QHE states was measured only in a 'two-terminal' geometry, $\kappa_{2T}T$. In such measurements[14, 15, 16, 17, 18], a single floating ohmic contact was used as a heat-current source with a known temperature (in analogy to a two-terminal electrical conductance measurement, where one terminal with a known voltage serves as a current source). However, the two-terminal thermal conductance, under some conditions, can be affected by multiple phenomena at the edge and thus differ from the topological $\kappa_{xy}$ even if the bulk is thermally insulating ($\kappa_{xx} = 0$)[18, 19]. The reason for this alleged contradiction is that the quantization of the two terminal thermal conductance ($\kappa_{2T} = |\kappa_{xy}|$) requires, in addition, the full thermalizing of counter-propagating (downstream and upstream) edge modes[20], if present.

The states whose edge is known to support topological counter-propagating modes are the Particle-Hole (P-H) Conjugated states[21, 22]. These fractional QHE states form when a Landau-level is more than half-full, most prominently at fillings $\nu = \frac{p}{2p-1}$, $p > 1$ integer, in the lowest spin-split Landau level[21, 22]. Charge equilibration between their counter-propagating edge modes happens fast (in less than $5\mu m$ in GaAs[19, 23]), allowing charge to flow only downstream and renders the upstream modes neutral[24, 25], thus guaranteeing the quantization of the two-terminal electrical conductance $G_{2T} = G_{xy}$ for the typical length-scale of most samples. However, inter-mode thermal equilibration is less efficient and depends on local microscopic details. As a result, for $\nu = 2/3$ (with charge $n_d = 1$ and neutral $n_u = 1$), the reported values of $\kappa_{2T}$ span from $0.3\kappa_0$ to $2\kappa_0$ (while $\kappa_{xy} = 0$)[15, 18, 19]. In a recent work, the thermal equilibration length of $\nu = 2/3$ was found to exceed $200 \mu m$ in GaAs samples[19].

To reliably measure $\kappa_{xy}$ of P-H Conjugated states, one must go beyond the two-terminal configuration and apply local measurements - say the temperature - separately for the downstream and upstream edge modes. Several local temperature measurement techniques had been reported in the QHE regime; those include quantum dot thermometry[22], partitioned noise thermometry[26, 27], and upstream (neutral) noise thermometry[19, 28]. However, none of these approaches offers a reliable, independent measurement of the temperatures of non-equilibrated counter-propagating modes. We adopted a different approach.

Here, we introduce a novel technique, which allows measuring the power carried by all types of edge modes, whether integer or fractional, propagating upstream or downstream. Moreover, we can measure the excess power carried by an edge mode, whether in equilibrium distribution with a definite temperature or a non-equilibrium distribution. This technique opens the path to perform multi-terminal thermal conductance measurements and to extract $\kappa_{xy}$ even when the edge modes are unequilibrated.

The key idea of this technique is the conversion of the energy carried by the edge modes to thermal energy of electrons in a floating small metallic reservoir, which serves as a power meter (PM). Once the PM's temperature is calibrated against a known heating power, one can determine the power absorbed in the PM by measuring its temperature $T_{PM}$. Importantly, This method relies only on the thermalization of electrons in a reservoir, and is thus applicable to any two dimensional topological material as long as ohmic contacts could be implemented along the edge.

In more details, when edge modes carrying power $P$ (which we wish to determine) impinge upon the PM and cause its temperature to increase, power will leave the heated PM in two ways, $P_{out} = P_e + P_\gamma$, with $P_e$ the power evacuated via QHE edge modes and $P_\gamma$ the power evacuated by lattice phonons. The temperature $T_{PM}$ is determined by the net power dissipated in the PM at equilibrium, $P = P_{out}(T_{PM})$. In our devices, we measure $T_{PM}$ by probing the Johnson-Nyquist (J-N) noise[29, 30] in a separate contact $A_{PM}$ (Fig. 1a). Calibration of the PM, i.e, converting the measured temperature to power, is accomplished with a separate measurement (Fig. 1b). We apply a known power at the PM by sourcing two opposite-sign DC currents, $I_{cal}$ and $-I_{cal}$, from two contacts $S_1^{\text{cal}}$ & $S_2^{\text{cal}}$, respectively, and measure the resulting PM's temperature. No direct current leaves the PM since its potential is zero. An equality $P_{\text{out}}(T_{\text{PM}}) = \frac{1}{G_{2\text{T}}} I_{\text{cal}}^2$ defines the equilibrium conditions for the calibration process.

The device, shown in Fig. 1c, consists of two floating small ohmic contacts (each with area $15 \times 2 \ \mu m^2$). Each contact is connected to three regions of the 2DEG, separated by narrow etched areas. The contact on the left serves as a 'temperature source' (S), and the contact on the right serves as a power-meter (PM). By sourcing opposite-sign DC currents, $I_S$ & $-I_S$, from contacts $S_1$ & $S_2$, respectively, we dissipate electrical power in S, causing its temperature $T_S$ to increase; yet, its potential is zero - ensuring that no current flows to PM and causes unwanted heating. Two metallic gates, 'gate-down' (GD) and 'gate-up' (GU), control whether the heated edge modes that emanate from S flow to the PM or the cold grounds.

In the experiment, $T_S$ and $T_{\text{PM}}$ are probed simultaneously by measuring the spectral density of the J-N noise by two amplifiers located downstream from S and PM (denoted as $A_S$ and $A_{\text{PM}}$ in Fig.1c). The measured low-frequency noise is proportional to the excess temperature of the heated contact from which the edge modes emanate (itis independent of the edge modes' temperature arriving at the amplifier contact). A separate calibration measurement allows deducting the impinging power on the PM. In a typical experiment, we measure the relation between the arriving power at the PM and the corresponding source temperature.

We start by measuring two particle-like states, the integer QHE state $\nu = 2$ and the fractional QHE state $\nu = 4/3$. In both, the edge hosts only two downstream edge modes. By changing the voltage on GD, we can allow either zero, one, or two heated downstream edge modes to flow from S to PM (the remaining edge modes flow to the grounded contact). For the measurement of these states, GU was kept open.

The measured power arriving at the PM is plotted as a function of the source temperature in Fig. 2 for two, one, and zero heated edge modes transmitted to the PM. We find a good agreement between the measured power and the theoretical expectation for the power carried by each heated mode, $P = \frac{\kappa_0}{2}(T_s^2 - T_0^2)$, validating our measurement technique. Unlike two-terminal measurements where the temperature must be low enough to avoid efficient heat evacuation to phonons, in this three-terminal measurement, there is no upper limit on $T_S$ since the phonon contribution is taken into account in the calibration process. Moreover, by opening both GD and GU, we can assure that there is no efficient 'parallel' heat conductance from S to PM (for example, due to a finite longitudinal thermal conductance $\kappa_{xx}$).

The advantage of the multi-terminal approach is particularly apparent in states that support counter-propagating modes, such as P-H conjugated states. It allows measuring independently the power carried by the downstream and upstream modes. For any P-H conjugated state, the total power carried by the downstream modes emanating from a contact with a fixed temperature $T_S$ after propagating a length $L$ is,

$$P_d(L) \equiv \frac{\kappa_d(L)}{2}(T_S^2 - T_0^2) = \frac{\kappa_0}{2} n_d(T_S^2 - T_0^2) - P_{d \to u}(L), \tag{1}$$

where $P_{d \to u}(L)$ is the power "backscattered" from the hot downstream edge modes to the cold upstream ones along the propagation distance $L$. Similarly, the total power carried by upstream modes (at the opposite edge of the sample) is,

$$P_u(L) \equiv \frac{\kappa_u(L)}{2}(T_S^2 - T_0^2) = \frac{\kappa_0}{2} n_u(T_S^2 - T_0^2) - P_{u \to d}(L), \tag{2}$$

where $P_{u \to d}$ is the "backscattered" power from the now hot upstream edge modes to the cold downstream (counter-propagating) modes.

As alluded above, the process of thermal equilibration among the counter-propagating edge modes is technically uncontrolled and not well understood. The reflected power back to the source can depend on microscopic parameters such as edge potential distribution, edge roughness, modes' velocities, temperature, and propagation length. However, under general conditions, the total backscattered power is equal in the downstream and the upstream sides of the mesa, namely, $P_{d \to u}(L) = P_{u \to d}(L)$ (see SI 5). One can thus extract $\kappa_{xy}$ by separately measuring the power carried by the downstream modes and the upstream modes for a fixed $L$,

$$P_d - P_u = \frac{(\kappa_d - \kappa_u)}{2}(T_S^2 - T_0^2) = \frac{\kappa_{xy}}{2}(T_S^2 - T_0^2). \tag{3}$$

We tuned the magnetic field to the ubiquitous P-H Conjugated state $\nu = 2/3$. This state supports (in the absence of edge-reconstruction) a single downstream charge mode and a single upstream neutral mode, corresponding to $\kappa_{xy} = 0$. By closing GD (GU) and opening GU (GD) we measure the power carried from S to PM by the downstream (upstream) side of the mesa - $P_d$ ($P_u$) (Fig. 3a-b). We extract the downstream (upstream) thermal conductance $\kappa_d$ ($\kappa_u$) by separately linear fitting the measured downstream (upstream) power arriving at the PM as a function of source temperature squared (Fig. 3c). Here, we find $\kappa_d = (0.43 \pm 0.03)\kappa_0$, $\kappa_u = $ $(0.39 \pm 0.03)\kappa_0$. Applying Eq. 3 gives $\kappa_{xy} = \kappa_d - \kappa_u = (0.04 \pm 0.03)\kappa_0$, which is very close to the expected value of $\kappa_{xy} = 0$. The two-terminal thermal conductance could also be extracted from the measurement as a sum of the downstream and the upstream thermal conductance coefficients, $\kappa_{2T} = \kappa_u + \kappa_d = 0.82 \pm 0.04\kappa_0$. We observe a non-equilibrated two-terminal thermal conductance, indicating that propagation length is much shorter than the thermal equilibration length. Nonetheless, we measure the thermal Hall conductance to be close to the topological value, demonstrating the universality of this approach for measuring $\kappa_{xy}$.

So far, we considered energy carried by edge modes with an equilibrium distribution (the excess power results from an elevated temperature of the modes). Our configuration allows measuring energy transport by edge modes with a non-equilibrium distribution. A simple example of a non-equilibrium distribution is a 'double-step distribution' (DSD)[31,32]. It can be formed by partitioning a biased edge mode in a quantum point contact (QPC) constriction. The distribution of the particles in the outgoing mode will be the statistical sum of the partitioned biased and unbiased incoming modes, $f = tf_V^{FD} + (1-t)f_0^{FD}$, where $f_V^{FD}$ ($f_0^{FD}$) is the Fermi-Dirac distribution of the particles in the biased (unbiased) mode, and $t$ is the QPC's transmission probability. This out-of-equilibrium distribution supports current fluctuations, with a low-frequency spectral density given by the well-known shot noise formula[33,34], $S_I(f \to 0) = 2e^\star It(1-t)\chi(\frac{e^\star I}{2k_b T G_{2T}})$, where $I$ is the impinging DC current, $e^\star$ the particles' charge and $\chi(x) = \coth(x) - \frac{1}{x}$. The excess power carried by the partitioned edge mode has two contributions: the trivial electrical power due to the transmitted direct current, $P_{DC} = \frac{1}{2G_{2T}}(tI)^2$, and the power stored in current fluctuations at all frequencies,

$$P_{AC} = \frac{1}{2G_{2T}} \int df S_I(f) = \frac{1}{2G_{2T}} I^2 t(1-t). \tag{4}$$

(See SI 6 for the derivation of Eq. 4 and further discussion). Interestingly, the quasi-particle charge $e^\star$ and the temperature $T$ drop out from the expression. The total power is $P = P_{DC} + P_{AC} = \frac{1}{2G_{2T}}tI^2$.

The device used to measure the energy carried by the DSD is similar to the one employed above (Figs. 4 a-d); however, we no longer use a heated source contact to excite the edge modes. Here, we source current $I$ from $S_I$, partition it by a partly pinched QPC (located at the center of the gate-down), forming a DSD that supports energetic fluctuations. The partitioned beam propagates for approximately $5 \mu m$ before being absorbed in the PM.

To measure the energy stored in the current fluctuations, we added a 'gate-arm', GA, which allows pinching off the region to the right of the PM. When GA is closed, current conservation dictates that the total low-frequency noise that flows to the amplifier is the partitioned shot-noise generated at the QPC (Fig. 4a). In this configuration, the heating of the PM will not cause any J-N noise (see SI 6). However, when GA is open, J-N noise will be generated on top of the shot noise due to the elevated temperature of the PM (Fig. 4b). By subtracting the shot noise, we are able to isolate the thermal fluctuations and determine $T_{\text{PM}}$ (see Methods). A calibration process (similar to the one described above) enables the extraction of the impinging power from $T_{\text{PM}}$. Here, we source to the PM unpartitioned current $I_{\text{cal}}$ from $S_{\text{I}}$ (by opening GD), and measure $T_{\text{PM}}$ as a function of the dissipating power $P = \frac{1}{4 G_{2\text{T}}} I_{\text{cal}}^2$ (Fig. 4c).

We concentrate on filling $\nu = 2$. At first, the DG was tuned to partition the outer edge mode while the inner one was fully reflected. From the total power measured by the PM, we subtract the dissipated DC power, thus isolating the power carried by the fluctuations $P_{\text{AC}}$ (see Methods). $P_{\text{AC}}$ is plotted as a function of the impinging current for four different transmission probabilities of the QPC (Fig. 4e). Indeed the measured power $P_{\text{AC}}$ agrees well with its expected value without any fitting parameters. This result is independent of any prior knowledge regarding $e^*$, the temperature, or even the gain of the amplifier (see SI 4).

However, when the GD was tuned to partition the inner edge mode, a smaller than predicted power (by as much as 40%) arrives at the PM (Fig. 4f). The missing energy of the partitioned inner edge mode of $\nu = 2$ is unexpected, as Eq. 4 is independent of the partitioned charge and the temperature. We note that our measurement technique is neither sensitive to charge redistribution along the edge[31] nor to energy transfer to the outer edge mode[35] (since PM is coupled to both modes). Three possible reasons for this discrepancy are: (i) The inner mode loses power to the bulk; (ii) The PM contact is not coupled well to the inner edge mode; (iii) power is lost at the QPC, possibly to excited non-topological neutral modes[36]. We note that the first two reasons are in contradiction with our observation of the correct $\kappa_{\text{xy}}$ of $\nu = 2$ (which implies that both edge modes carry the expected power when thermally biased). More research is required to understand this discrepancy, and its relation to other unexplained observations regarding innermost edge modes, such as vanishing interference and non-Gaussian noise[37].

We developed a new method that enables local power measurement of 1D edge modes. We demonstrate the robustness of this realization in the QHE regime by measuring the power carried by integer and fractional edge modes. It is possible to separately measure the power carried by the upstream and the downstream modes by controlling the device geometry. This method allows overcoming the difficulty of the traditional two-terminal measurements that depend on equilibration between the counter-propagating modes. The power of this method is exemplified by measuring the topological thermal Hall conductance of $\nu = \frac{2}{3}$, while the edge modes are far from being thermally equilibrated (as the propagation length is much shorter than the thermal equilibration length). In this case, we find $\kappa_{xy} = 0.04 \pm 0.03 \kappa_0$, which is in very close agreement with the theoretical understanding of the topological order of this state. We further demonstrate the ability to measure the power carried by an out-of-equilibrium double-step energy distribution of partitioned edge modes. This new method allows a more holistic study of energy transport in the QHE regime, and could be implemented to other 2D topological materials.

## Author contribution:

R.A.M and S.K fabricated the devices, performed the measurements and analyzed the data. M.H supervised the experiment and the analysis. V.U grew the GaAs heterostructures. All authors contributed to the writing of the manuscript.

## Acknowledgements

We acknowledge Christian Spånslätt, Jinhong Park, Alexander D. Mirlin and Kyrylo Snizhko for usefull discussions. M.H. acknowledges the continuous support of the Sub-Micron Center staff, the support of the European Research Council under the European Community's Seventh Framework Program (FP7/2007-2013)/ERC under grant agreement number 713351, the partial support of the Minerva foundation under grant number 713534.

## Methods:

**Sample preparation:**

The Ohmic contacts and gates were patterned using standard e-beam lithography-liftoff techniques. The Ohmic contact consists, from the surface and upwards, Ni (7nm), Au (200nm), Ge (100nm), Ni (75nm), Au (15nm) alloyed at $440°C$ for 50 seconds. After the preparation of the ohmic contacts, the sample was covered by 20nm $\text{HfO}_2$ deposited at $200°C$. The ohmic contact, $\text{HfO}_2$ and the gate evaporated on top forms an 'on-chip' capacitor with a capacitance of $\sim 0.5 pF$. In order to wire-bond the sample, we etched the $\text{HfO}_2$ (from the bonding pads) using a *buffered oxide etch*. The gate electrode consists of 5nm Ti and 15nm Au.

**Noise calculation - multi-terminal thermal conductance:**

In the thermal conductance measurements, the noise measured in the PM amplifier is associated with the temperature of the power meter. However, there is a noise contribution that originates from the elevated temperature of the source and must be subtracted. We consider the general case where the source is at a temperature $T_S$ the power meter is heated to a temperature $T_{\text{PM}}$. The two-terminal electrical conductance from the source to the PM (ground) is $G_{\text{S}\to\text{PM}}$ ($G_{\text{S}\to\text{G}}$). The spectral density of the current fluctuations flowing to the power meter is given by,

$$S_{\text{S}\to\text{PM}} = 2k_\text{b}\alpha \frac{G_{\text{S}\to\text{PM}} G_{\text{S}\to\text{G}}}{G_{\text{S}\to\text{PM}} + G_{\text{S}\to\text{G}}} (T_\text{S} - T_0), \tag{M1}$$

where $\alpha$ is a pre-factor that accounts for smaller noise in P-H conjugated states with unequilibrated edge modes[19]. For the measured states, we had $\alpha = 1$ for $\nu = 2$ and $\nu = \frac{4}{3}$ and $\alpha = \frac{3}{4}$ for $\nu = \frac{2}{3}$. The excess fluctuations are divided at the PM contact between all the arms. As a result, the current fluctuations arriving at the $A_{PM}$ have two uncorrelated contributions,

$$S_{\text{PM}} = \left(\frac{G_{\text{PM}\to\text{A}}}{G_{\text{PM}\to\text{G}} + G_{\text{PM}\to\text{A}}}\right)^2 S_{\text{S}\to\text{PM}} + 2k_\text{b}\alpha \frac{G_{\text{PM}\to\text{A}} G_{\text{PM}\to\text{G}}}{G_{\text{PM}\to\text{A}} + G_{\text{PM}\to\text{G}}} (T_{\text{PM}} - T_0). \tag{M2}$$

Here $G_{\text{PM}\to\text{A}}$ ($G_{\text{PM}\to\text{G}}$) is the conductance from the PM to $A_{\text{PM}}$ (ground). As the temperature of the source is separately measured by means of noise in $A_\text{S}$, the first term is known. Eq. M2 allows extracting $T_{\text{PM}}$ from the noise picked up at $A_{\text{PM}}$. See SI 3 for a 'step-by-step' analysis of the raw data.

**Noise calculations - double-step distribution:**

The noise measured at the amplifier has two contributions, the first is the J-N noise emanated from the PM contact due to its elevated temperature and the second term is the shot noise generated at the QPC - $S_{\text{shot}}$. Similarily to Eq. M2 we can write the total noise,

$$S_{\text{PM}} = \left(\frac{G_{\text{PM}\to\text{A}}}{G_{\text{PM}\to\text{G}} + G_{\text{PM}\to\text{A}}}\right)^2 S_{\text{shot}} + 2k_\text{b} \frac{G_{\text{PM}\to\text{A}} G_{\text{PM}\to\text{G}}}{G_{\text{PM}\to\text{A}} + G_{\text{PM}\to\text{G}}} (T_{\text{PM}} - T_0). \tag{M3}$$

Since we measured $\nu = 2$ and our device had two arms, we can write it in a compact form as,

$$S_{\text{PM}} = \frac{1}{4}S_{\text{shot}} + k_{\text{b}}G_{2\text{T}}(T_{\text{PM}} - T_0), \tag{M4}$$

with $G_{2\text{T}} = \frac{2e^2}{h}$. In this experiment, we used only one amplifier, thus $S_{\text{shot}}$ could not be simultaneously measured. We measured the shot noise separately by closing the gate-arm (see Fig 4a). In this configuration, the noise arriving at the amplifier is the partitioned shot noise generated at the QPC (see SI 4).

**Power carried by DC currents:**

Considering a 'double-step distribution', the total power dissipated at the PM has two contributions: the power carried by the fluctuations at all frequencies $P_{\text{AC}}$, and the dissipated part of the power carried by the DC currents. The second contribution is subtracted from the measured power in order to extract $P_{\text{AC}}$.

When an edge mode carrying a current $I$ impinges upon a floating contact it does not dissipate its full power $P_{\text{DC}}^{\text{in}} = \frac{1}{2G_0}I^2$ on the contact. One has to subtract the power carried by the DC currents in the outgoing edges,

$$P_{\text{DC}}^{\text{dis}} = P_{\text{DC}}^{\text{in}} - \frac{N}{2}V^2 G_0, \tag{M5}$$

with $V$ the contact's potential and $N$ the number of outgoing edge modes (here, $N = 4$). When we partitioned the outer edge of $\nu = 2$ with transmission $t$, the dissipated power is $P_{\text{DC}}^{\text{dis}} = \frac{3}{8G_0}(tI)^2$. When the inner mode is partiotioned $P_{\text{DC}}^{\text{dis}} = \frac{I^2}{8G_0}(3 + 3t^2 - 2t)$. See SI 4.4 for further discussion.

# Figures:

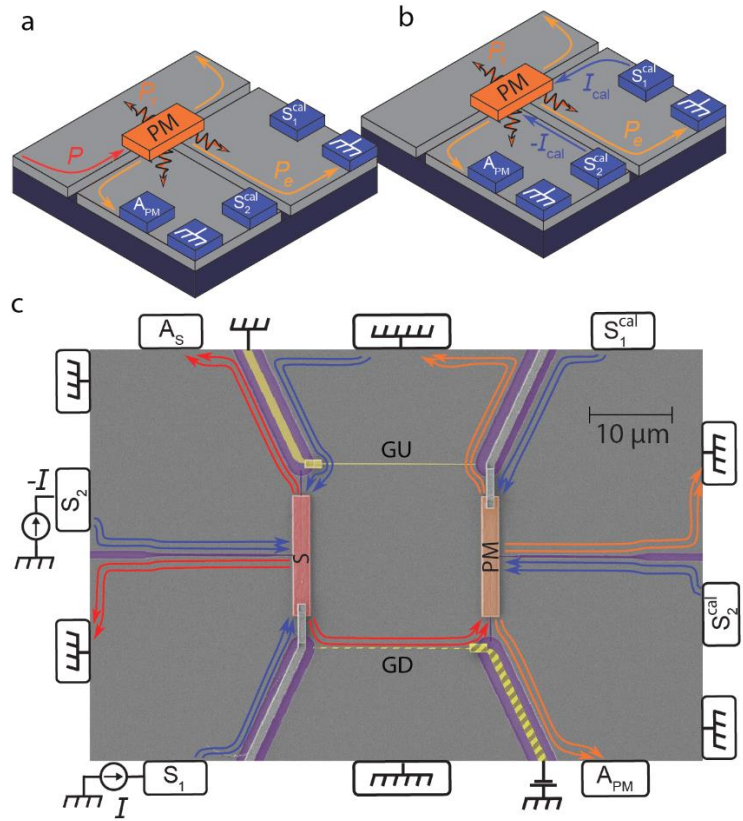

**Figure 1. Device configuration for measuring heat transfer** – (**a,b**) Measurement configuration of the power-meter (PM). The PM is a floating small ohmic contact connected to three regions of 2DEG. (**a**) Measuring the impinging power on the PM. A heated edge mode (*P*) leaving a heat-source (not seen) impinges on the PM, increasing its temperature, $T_{PM}$. The resultant Johnson-Nyquist noise is measured by the amplifier connected at $A_{PM}$. In equilibrium, the power evacuated from PM (via phonons $P_\gamma$ and via edge modes $P_e$) is equal to the impinging power $P$. (**b**) Calibration of the PM. $T_{PM}$ is measured for a known impinging power carried by two DC currents $I_{cal}$ and $-I_{cal}$, sourced simultaneously by $S_1^{cal}$ and $S_2^{cal}$, with $P_{cal} = \frac{I_{cal}^2}{G_{2T}}$. This calibration allows converting $T_{PM}$ to the impinging power $P$. (**c**) False colors SEM image of the tested structure. The mesa (gray) is divided into several parts by etched regions (purple). The mesa parts are connected by the floating metallic ohmic contacts, which serve as temperature sources (S, colored red) and a power meter (PM, colored orange). Both contacts' capacitance to the ground is enhanced by an insulated top plate ($\sim 0.5 pF$, grounded via gray leads). The arrows describe two downstream propagating edge modes corresponding to the edge structure of $\nu = 2$. When currents $+I$ and $-I$ are sourced from $S_1$ and $S_2$, respectively, the source S heats up to temperature $T_S$, measured via J-N noise that is picked up at the source amplifier at $A_S$. In the shown configuration, gate-down (GD, colored dashed yellow) is closed (by applying a negative voltage), forcing the heated edge modes to flow from S to PM. Gate-up (GU, colored full yellow) is open, thus grounding the upstream modes (if there). The edge modes heat the PM to temperature $T_{PM}$,

measured by noise at the PM amplifier $A_\mathrm{PM}$. $T_\mathrm{PM}$ is later converted to power using a calibration measurement (see Fig. 1**b**).

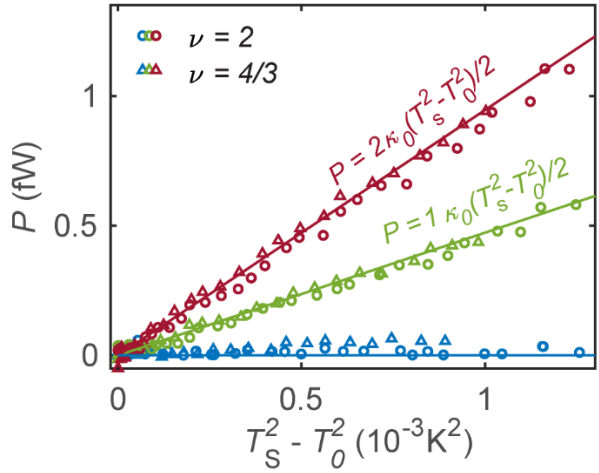

**Figure 2. Power carried by downstream modes** - Power $P$ carried by the downstream edge modes of $\nu = 2$ (circles) and $\nu = 4/3$ (triangles) as a function of the source temperature $T_S$. The different colored markers correspond to a different number of heated edge modes emanating from $S$ and transmitted across the partly closed GD (by applying a negative voltage to the gate) to arrive at PM. Red - two modes; Green - one mode (inner mode); Blue - zero modes (the small observed power in $\nu = 4/3$ corresponds to a finite $\kappa_{xx}$). For both integer and fractional states, there is a good agreement with the prediction of $P = n_d \frac{\kappa_0}{2}\left(T_S^2 - T_0^2\right)$, without any fitting parameters (colored lines).

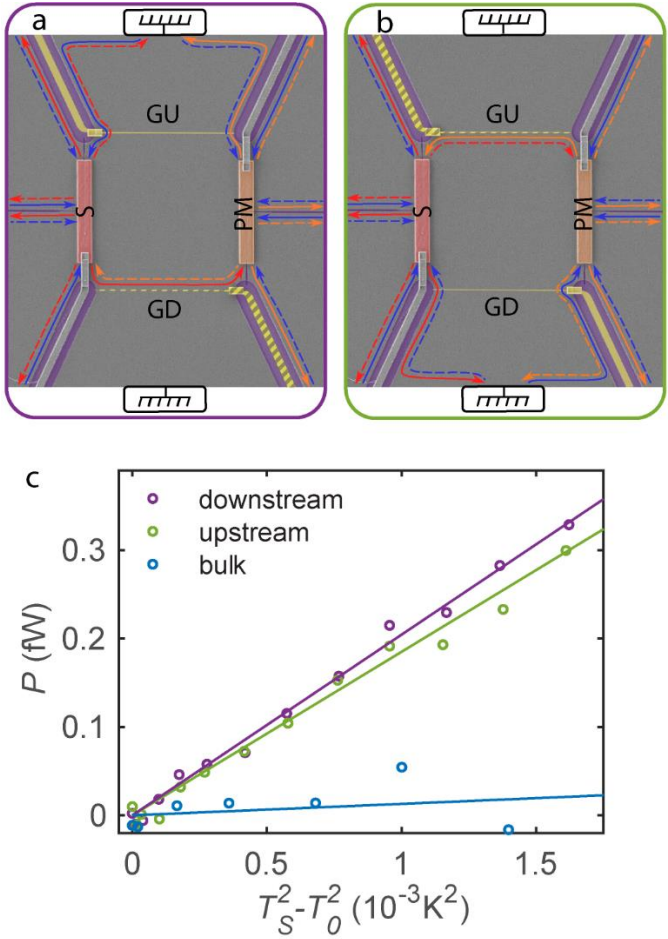

**Figure 3. Thermal conductance of the Particle-Hole Conjugated $\nu = 2/3$ state** – (**a**) Device configurations used to measure the power carried by the downstream mode. (**b**). Similar configuration to measure the power carried by the upstream mode. The arrows mark the edge modes of the $\nu = 2/3$ state – full (dashed) arrow corresponds to downstream charged mode (upstream neutral mode). In (**a**) GD is closed (negative voltage applied on the gate), and GU is open. Excess power carried to PM by the heated downstream mode. In (**b**) GU is closed and GD is open. Only the upstream mode carries power to PM. (**c**) Downstream power $P_d$ (purple markers) and upstream power $P_u$ (green markers) measured in the configuration depicted in (**a**) and (**b**), respectively. Linear fitting $P_d$ and $P_u$ versus source temperature $T_S^2$, with fittings $\kappa_d/\kappa_0 = 0.43 \pm 0.03$ and $\kappa_u/\kappa_0 = 0.39 \pm 0.03$ ; and consequently, $\kappa_{xy} = \kappa_d - \kappa_u = (0.04 \pm 0.03)\kappa_0$, with the latter agreeing nicely with the expected $\kappa_{xy} = 0$. The blue markers, linearly fitted with the blue line, were measured when both GD and GU were open. Here, the minute power arriving in the PM is either due to heat propagating through the bulk or due to slight reflection from the 'open' gates.

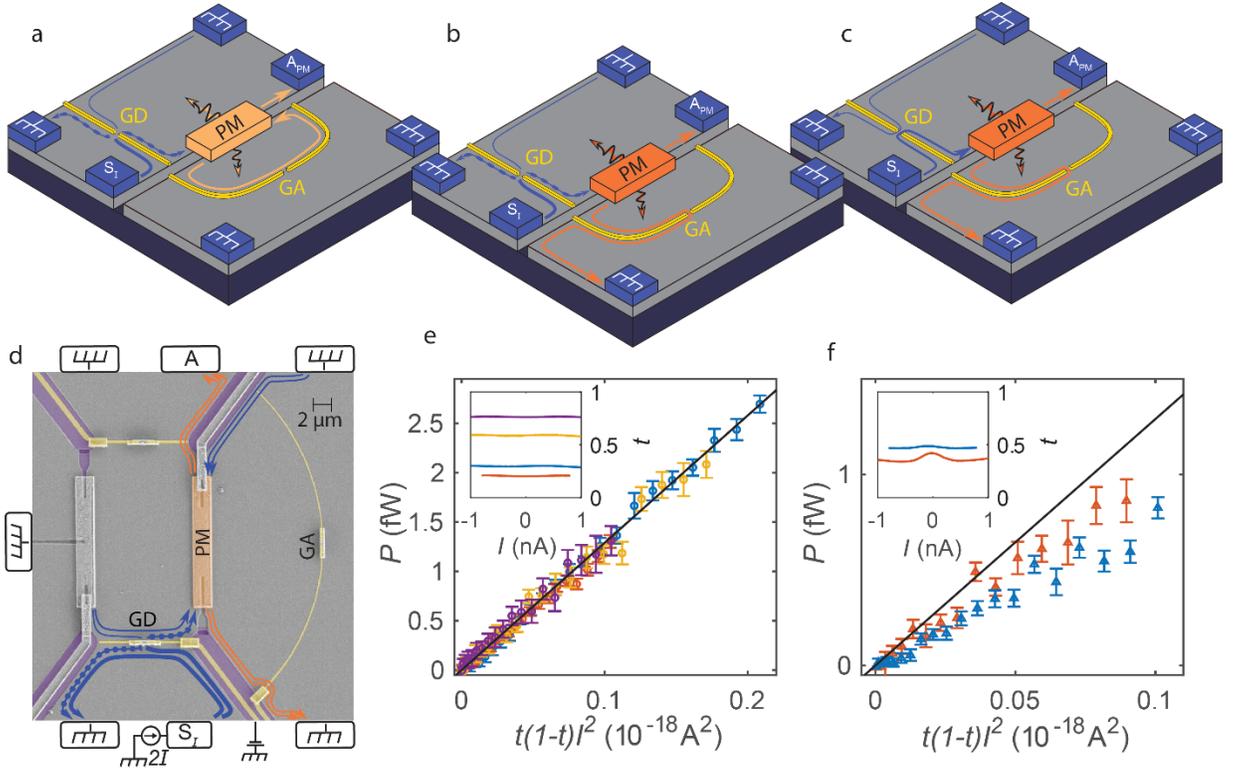

**Figure 4. Measuring power carried by out-of-equilibrium edge mode distributions at $\nu = 2$. (a-c)** Schematics of the three measured gate configurations **(a)** Configuration for shot noise measurement. A biased edge mode (thick line) is partially transmitted through a 'split gate' QPC located at the center of gate-down (GD), forming a 'double-step distribution' (DSD). The associated shot noise passes through the PM to the amplifier $A_{PM}$. Since the 'arm gate' (GA) is closed, no added low-frequency fluctuations are added. **(b)** Configuration to measure the power carried by the DSD. By opening GA we enable the J-N noise, which is excited due to the PM heating, to be added to the shot noise. The noise is picked up at $A_{PM}$. This noise is used to extract the PM's temperature, $T_{PM}$ (see Methods). **(c)** Set up for calibration measurement. The GD is unbiased (open), and the biased edge modes are fully transmitted toward the PM, heating it by a known dissipated power $P = \frac{I^2}{4G_{2T}}$. Measuring $T_{PM}$ against the known power manifests the calibration process and enables to infer absorbed power from the measured temperature. **(d)** False-color SEM image of the device used to measure the power carried by the DSD. The power meter (PM, colored orange) is connected to two regions of 2DEG, otherwise separated by etched regions (purple). The lines represent the edge modes of $\nu = 2$ (thin-unbiased; thick-biased; broken-partitioned beam). **(e,f)** AC power is carried by the outer edge mode **(e)** and the inner edge mode **(f)** DSDs. The power is plotted against the impinging current squared multiplied by $t(1-t)$, and the black line with slope $\frac{h}{2e^2}$, is expected according to Eq. 4. The different colored markers correspond to various transmission coefficients of the QPC. The non-linear differential transmission is shown as a function of the current in the inset of (**e**) and (**f**). We find that the power carried by DSD of the outer edge mode agrees well with the expectations without any fitting parameter (**e**). On the other hand, we measured a lower power **(f)** when the inner mode is partitioned.

**Supplemantary Information for: Direct Determination of the Topological Thermal Conductance via Local Power Measurement**

Ron Aharon Melcer[1], Sofia Konyzheva[1,2], Moty Heiblum[1], and Vladimir Umansky[1].

1. *Braun Center for Submicron Research, Department of Condensed Matter Physics, Weizmann Institute of Science, Rehovot 761001, Israel*
2. *Laboratory of Quantum Materials (QMAT), Institute of Materials (IMX), École polytechnique fédérale de Lausanne, 1015 Lausanne, Switzerland*

# 1 REPRODUCIBILITY

We measured four samples fabricated from three different GaAs MBE grown heterostructures. The chip design of all is similar to the one shown in Fig. 1c. We tested two variations: (i) The length of edge propagation from source to power-meter changed from $20\mu m$ to $10\mu m$; (ii) The number of 2DEG regions connected to PM was changed from three to two. In addition, we made several measurements in a higher base temperature $T_0$. The results are consistent in all measured devices and geometries. We summarized all the measured results in Table S1.

# 2 SHOT NOISE

In addition to measuring the power carried by the fluctuations associated with the 'double-step' distribution, we also measured the low-frequency fluctuations, i.e shot noise. We sourced current $I$ from $S_I$ and partitioned the edge mode in a quantum point contact (QPC) located at the center of 'gate-down' (GD). By closing 'gate-arm' (GA) we can measure the shot noise in the amplifier contact (see Fig. 4a). Assuming independent tunneling events of quasiparticles with charge $e^*$, the shot noise is given by,

$$S_{\text{shot}} = 2e^\star I t(1-t)\left(\coth\left(\frac{e^\star V}{2k_b T}\right) - \frac{2k_b T}{e^\star V}\right), \quad (S1)$$

with $t$ the transmission probability, $k_b$ Boltzmann constant, $T$ the temperature and $V = I/G$ the mode's voltage ($G$ is the edge mode's conductance). Fitting the measured noise to formula S1 allows the extraction of the quasiparticle charge $e^\star$ [S1, S2] and the temperature [S3]. For integer filling edge modes with $G = G_0 = \frac{e^2}{h}$ ($e$ the electron's charge and $h$ plank's constant), one expects $e^\star = e$. As shown in Fig. S1a, all the measurements of the noise in the outer edge mode are consistent with Eq. S1.

Gain calibration was done via shot noise in the outer edge mode of filling factor 2. Once the gain is fixed, we fitted the charge and the temperature. In all measured transmissions of the outer mode, we find $e^\star = e$ and $T \sim 20 mK$ (which is consistent with the measured cryostat temperature of $15 mK$). See exact fitting coefficients in the caption of Fig. S1a.

The results for the shot noise of the partitioned inner edge mode are plotted in Fig. S1b. Here, we find a smaller noise than expected. By fitting the noise to Eq. S1 we find $e^\star/e = 0.65 \pm 0.01$ ($e^\star/e = 0.55 \pm 0.02$) and $T = 7.0 \pm 0.5 mK$ ($T = 17 \pm 2 mK$) for QPC transmission of $t = 0.47$ ($t = 0.38$). These results remain unexplained.

## 3 MEASUREMENT PROTOCOL – MULTI-TERMINAL THERMAL CONDUCTANCE:

To better clarify our measurement technique, we show in this section a 'step by step' analysis of the raw data. As an example, we consider the data for $\nu = 2$ shown in Fig. 2.

### 3.1 NOISE ACQUISITION:

We heat the Source (S) by sourcing currents $I$ and $-I$ from $S_1$ and $S_2$, respectively (see Fig. 1c). The strength of the low-frequency current fluctuations is measured in two separate contacts $A_S$ and $A_{PM}$, located downstream from S and the power-meter (PM), respectively. We repeat this measurement for three different voltages on 'gate-down' (GD), corresponding to two, one, and zero fully transmitted edge modes (from S to PM). The raw data is plotted in Fig. S2a.

### 3.2 EXTRACTING TEMPERATURES:

The elevated temperature of S is associated with Johnson-Nyquist (J-N) noise. The low-frequency power spectral density of the current fluctuations flowing to $A_S$ is given by

$$S^{JN}_{S \to A_S} = 2k_b \alpha G^\star_{S \to A_S}(T_S - T_0). \tag{S2}$$

where $k_b$ is the Boltzmann constant, $T_S$ ($T_0$) is the source's temperature (base temperature), and $\alpha$ is a pre-factor that accounts for a smaller noise in particle-hole conjugated states with un-equilibrated edge modes [S4] (for $\nu = 2$, $\alpha = 1$). $G^\star_{S \to A_S}$ is the effective conductance related to the device geometry. It is given by

$$G^\star_{S \to A_S} = \frac{G_{S \to A_S}(G_{S \to PM} + G_{S \to G})}{G_{S \to A_S} + G_{S \to PM} + G_{S \to G}}. \tag{S3}$$

with $G_{S \to A_S}$ the conductance from S to $A_s$, $G_{S \to PM}$ the conductance from S to the PM, and $G_{S \to G}$ is the conductance from S to the ground (G). For $\nu = 2$, we have $G^\star_{S \to A_S} = \frac{4}{3}G_0$ ($G_0 = \frac{e^2}{h}$, with $e$ the electron's charge and $h$ plank's constant) [S5].

The noise measured at $A_{PM}$ has two uncorrelated contributions. The first is the result of the elevated temperature of the PM. The power spectral density of the J-N noise is given by.

$$S_{PM \to A_{PM}}^{JN} = 2k_b \alpha G_{PM \to A_{PM}}^\star (T_{PM} - T_0), \quad (S4)$$

with,

$$G_{PM \to A_{PM}}^\star = \frac{G_{PM \to A_{PM}} G_{PM \to G}}{G_{PM \to G} + G_{PM \to A_{PM}}}. \quad (S5)$$

For $\nu = 2$, when the PM is connected to three arms $G_{PM \to A_{PM}}^\star = \frac{4}{3} G_0$.

The second contribution to the noise measured in $A_{PM}$ is due to low-frequency J-N current noise flowing from S to PM.

$$S_{S \to PM}^{JN} = 2k_b \alpha G_{S \to PM}^\star (T_S - T_0), \quad (S6)$$

with,

$$G_{S \to PM}^\star = \frac{G_{S \to PM}(G_{S \to G} + G_{S \to A_S})}{G_{S \to PM} + G_{S \to G} + G_{S \to A_S}}. \quad (S7)$$

$G_{S \to PM}^\star$ depends on the number of conducting modes transmitted (through GD) from S to PM. Hence, $G_{S \to PM}^\star / G_0 = \frac{4}{3}, \frac{5}{6}, 0$ for two, one and zero fully transmitted channels, respectively. These current fluctuations are equally divided in the PM between all the outgoing edge modes. We find that the total noise arriving to $A_{PM}$ is given by,

$$S_{A_{PM}} = 2k_b \alpha \left( G_{PM \to A_{PM}}^\star (T_{PM} - T_0) + \left( \frac{G_{PM \to A_{PM}}}{G_{PM \to G} + G_{PM \to A_{PM}}} \right)^2 G_{S \to PM}^\star (T_S - T_0) \right). \quad (S8)$$

Simultaneously measuring the noise in $A_S$ and $A_{PM}$ allows us to extract $T_S$ and $T_{PM}$ using Eq. S2 and S8. We plot $T_{PM}$ as a function of $T_S$ in Fig. S2b.

### 3.3 CALIBRATING THE PM:

We are interested in the power arriving at PM rather than its temperature. We accomplish this by performing a separate calibration measurement where we heat the PM directly with a known power (S remains unheated). We open both GD and GU and inject DC currents $I_{cal}$ & $-I_{cal}$ from $S_1^{cal}$ and $S_2^{cal}$, respectively. This leads to power dissipation $P = \frac{I_{cal}^2}{G_{2T}}$ ($G_{2T}$ is the state's two-terminal conductance). We measure the noise in $A_{PM}$ (Fig S2c), which, in this case, is proportional only to the PM's temperature according to Eq. S4 (Fig. S2d). Using the calibration process, we can plot the impinging power on the PM as a function of the source's temperature (Fig. S1e).

# 4 MEASUREMENT PROTOCOL – THE POWER OF 'DOUBLE STEP' DISTRIBUTIONS

In this chapter, we show a systematic analysis of the raw data of the power measurement of the 'double-step distribution' (DSD). As an example, we consider the outer edge mode of $\nu = 2$, shown in Fig. 4e.

## 4.1 SHOT NOISE MEASUREMENT

The first step is to measure the shot noise $S_{\text{shot}}$, i.e, the low-frequency spectral density of the fluctuations associated with the DSD. The DSD is generated by sourcing current $2I$ from the current source $S_I$ (in $\nu = 2$ a current $I$ will flow in each of the edge modes). The edge mode is partitioned in a QPC located at the 'gate-down' (GD) center. By closing the 'gate-arm' (GA), we 'deactivate' the power-meter (PM), not allowing it to add any low-frequency fluctuations on top of the shot noise (Fig. 4a of the main text). Important to note that the PM does heat up due to the impinging power. Nonetheless, since it is floating, its elevated temperature will not be associated with added current fluctuations in the outgoing edge modes (due to 'current conservation' of the arriving current fluctuations). We plot $S_{\text{shot}}$ vs. $I$ in Fig. S3a. The analysis of the noise is discussed in the dedicated section above.

## 4.2 THERMAL NOISE MEASUREMENT

By opening GA we allow J-N noise to emanate from the PM and arrive at the amplifier contact (Fig. 4b). The noise in this configuration is plotted against $I$ in Fig. S3b. The spectral density of the J-N noise is given by equations S4, S5 (notice that here there are only two arms),

$$S_{\text{PM} \to A_{\text{PM}}}^{\text{JN}} = 2k_b G_0 (T_{\text{PM}} - T_0) \, , \tag{S9}$$

The total noise arriving at $A_{\text{PM}}$ consists of the J-N noise and the shot noise that is divided at the PM contact (the fluctuations are uncorrelated),

$$S = S_{\text{PM} \to A_{\text{PM}}}^{\text{JN}} + \frac{1}{4} S_{\text{shot}} \, . \tag{S10}$$

Using Eq. S9 & Eq. S10 with the known $S_{\text{shot}}$ we extract $T_{\text{PM}}$. $T_{\text{PM}}$ is plotted against $I^2$ in Fig. S3c.

## 4.3 CALIBRATING THE PM

By opening GD, we can source DC current $I_{\text{cal}}$ from $S_I$ to the PM (Fig. 4c). The excess current leads to dissipation of power $P = \frac{1}{8G_0} I_{\text{cal}}^2$ in the PM, with J-N noise flowing to the amplifier contact (Eq. S9 & Fig. S2d). We plot the dissipated power vs. $T_{\text{PM}}$ in Fig. S2e, which defines the calibration process.

## 4.4 SUBTRACTING THE DC POWER

By combining the DSD noise measurement with the calibration measurement, we plot the dissipated power of the DSD vs. $I^2$ (colored markers in Fig. S2f). This power has two contributions, the power carried by the fluctuations $P_{AC}$ and the dissipated part of the power carried by the DC current. Generally, an edge mode carrying a DC current $I$ does not dissipate the full excess power $P_{DC}^{in} = \frac{I^2}{2G_0}$ in a floating contact. From the impinging power, one has to subtract the power that is carried by the outgoing DC currents $P_{DC}^{out} = \frac{N}{2} V_{PM}^2 G_0$, where $N$ is the number of integer modes leaving the contact (in our sample $N = 4$ since the PM connected to two arms at $\nu = 2$). The contact's potential $V_{PM}$ is determined by the sum of the impinging currents in all the incoming modes,

$$V_{PM} = \sum_i \frac{I_i}{NG_0}, \tag{S11}$$

with $I_i$ the current in the $i$'th incoming mode. The dissipated power $P_{DC}^{dis} = P_{DC}^{in} - P_{DC}^{out}$ could be written in terms of the incoming currents as

$$P_{DC}^{dis} = \frac{1}{2G_0}\left(\sum_i I_i^2 - \frac{1}{N}\left(\sum_i I_i\right)^2\right). \tag{S12}$$

When we partition the outer mode in our experiment, it is the only biased mode impinging at the PM. It carries DC current $tI$ ($t$ the QPC transmission probability), giving $P_{DC}^{dis} = \frac{3}{8G_0}(tI)^2$ (dashed lines in Fig. S2f). When the inner edge mode is partitioned, it carries current $tI$, with an un-partitioned outer mode with current $I$; thus, the total dissipated power is $P_{DC}^{dis} = \frac{I^2}{8G_0}(3 + 3t^2 - 2t)$. $P_{AC}$ is extracted by subtracting the dissipated DC power from the measured power $P_{AC} = P - P_{DC}^{dis}$ (Fig. S2g).

## 4.5 POWER MEASUREMENT IS INDEPENDENT OF GAIN CALIBRATION

As an aside, we wish to point out that the protocol described above allows the extraction of the power without the need to calibrate the amplification gain or the base temperature. The reason is that the measured temperature of the PM is only used as an intermediate result to compare the impinging powers of the DSD and the calibration measurement. Thus, it is sufficient to measure the increase of temperature $\Delta T = T_{PM} - T_0$ up to a multiplying constant, which eliminates the need to calibrate the gain. In the shown data, the amplification gain and $T_0$ were extracted from the shot noise measurements of the outer edge mode, described in the designated section above.

# 5 THERMAL HALL CONDUCTANCE

In this section, we wish to further discuss the physics of heat transport along the edge modes and the necessary conditions required for extracting $\kappa_{xy}$ from our power measurements. In order to do so, we consider the heat balance equations of the power-meter (PM) for the two geometries involved in the experiment. We will discuss a general quantum Hall effect (QHE) state with $n_\text{d}$ downstream modes and $n_\text{u}$ upstream modes.

First, we consider the configuration in which 'gate-down' (GD) is closed and 'gate-up' (GU) is open (Fig. 3a). The total excess power impinging upon the PM is,

$$P_\text{d} = \frac{\kappa_0}{2}\left(n_\text{d}\left(T_\text{d}^2(l_\text{d}) - T_0^2\right) + n_\text{u}(T_\text{u}^2(l'_\text{u}) - T_0^2)\right), \qquad (S13)$$

where $\kappa_0 = \frac{\pi^2 k_b^2}{3h}$ is the thermal conductance quantum; $T_\text{d}(l_\text{d})$ is the temperature of the downstream modes emanating from the source (S) after propagating the distance $l_\text{d}$ to the PM; $T_\text{u}(l'_\text{u})$ is the temperature of the upstream modes emanating from the ground (G) after propagating the distance $l'_\text{u}$ to the PM. We take the ground's temperature to be $T_0$. In Eq. S13 we have already included the assumption that there is no other thermal conduction mechanism from S to PM besides the QHE edge modes (we verify this assumption by performing a measurement where both GD and GU are open).

A second assumption we wish to implement is that no heat is lost from the edge to the environment along the propagation length from S to PM. This is a reasonable assumption as the length from S to PM is $20 \mu m$ or less in all the measured samples. Recent experiments suggest that both upstream and downstream modes are capable of carrying heat for such distances without significant dissipation [S4, S6]. This assumption implies that a heated mode can only lose its power to a counter-propagating mode on the same edge (via thermal equilibration). It allows us to rewrite the impinging power as

$$P_\text{d} = \frac{\kappa_0}{2}n_\text{d}(T_S^2 - T_0^2) - P_{\text{d}\to\text{u}}(T_S, T_\text{PM}, l_\text{d}) + P_{\text{d}\to\text{u}}(T_\text{PM}, T_0, l'_\text{u}), \qquad (S14)$$

where $T_S$ is the source's temperature, and the function $P_{\text{d}\to\text{u}}(T_1, T_2, l)$ describes the total power backscattered from the hot downstream modes emanating from contact with temperature $T_1$ to the upstream modes emanating from contact with temperature $T_2$ along the propagation length $l$.

There are two mechanisms contributing to this process of edge equilibration. The first is the backscattering of plasmon modes at the edge-mode contact interface [S4, S7]. This term is length independent and proportional to the difference in temperatures squared, $P^1_{\text{d}\to\text{u}}(T_1, T_2, l) \propto T_1^2 - T_2^2$. The exact coefficient depends on an exact theory of the edge modes. For example, for $\nu = \frac{2}{3}$ in the strong interaction fixed point [S4] $P^1_{\text{d}\to\text{u}}(T_1, T_2, l) = \frac{1}{2}\frac{\kappa_0}{2}(T_1^2 - T_2^2)$.

The second mechanism is inelastic scattering along the edge [S8], causing heat to flow from the hot modes to the cold ones. The details of this interaction are unknown, and the equilibration's efficiency can depend on microscopic details of the edge. It is convenient to describe the equilibration process via Newton's law of cooling [S9, S10], which introduces a thermal equilibration length $l_\text{eq}$ and implies,

$$P^2_{d\to u}(T_1, T_2, l) = f_{d\to u}\left(\frac{l}{l_{eq}}\right)\frac{\kappa_0}{2}(T_1^2 - T_2^2), \tag{S15}$$

with $f_{d\to u}$ describing the equilibration process along the edge. For $\nu = \frac{2}{3}$, $f_{d\to u}(x) = \frac{x}{1+x}$, and for a general P-H conjugated states with $n_u \neq n_d$

$$f_{d\to u}(x) = \frac{n_d n_u \left(\exp\left(\frac{x}{n}\right) - \exp\left(-\frac{x}{n}\right)\right)}{n_d \exp\left(\frac{x}{n}\right) - n_u \exp\left(-\frac{x}{n}\right)}, \tag{S16}$$

with $n = \frac{n_d n_u}{n_d - n_u}$. In Eq. S16 we assumed that the efficiency of equilibration is identical between all the modes along the edge. Two interesting limits are $f_{d\to u}(x=0) = 0$ and $f_{d\to u}(x\to\infty) = \min(n_d, n_u)$. Moreover, $f_{d\to u}$ is invariant under $n_d \leftrightarrow n_u$.

Even though Eq. 3 of the main text holds for a more general class of inter-mode interaction; we will adopt Newton's law of cooling mechanism in the remaining of this chapter. The main reason (other than its simplicity) is the experimental fact that in the temperature regime we measured, this model appears to describe our results well (as is evident by the linear dependence of the power with $T_S^2$). Describing the inter-edge interaction in this formalism allows writing the total backscattered power as,

$$P_{d\to u}(T_1, T_2, l) = \frac{\kappa_0}{2} g_{d\to u}\left(\frac{l}{l_{eq}}\right)(T_1^2 - T_2^2), \tag{S17}$$

where the function $g_{d\to u}$ includes both equilibration mechanisms. Under these definitions, the excess power impinging upon the PM is given by,

$$P_d = \frac{\kappa_0}{2}\left(\left(n_d - g_{d\to u}\left(\frac{l_d}{l_{eq}}\right)\right)(T_S^2 - T_0^2) + g_{d\to u}\left(\frac{l'_u}{l_{eq}}\right)(T_{PM}^2 - T_0^2)\right). \tag{S18}$$

The situation where GD is open and GU closed is very similar. Under the same assumptions, we can write the excess power arriving at the PM as,

$$P_u = \frac{\kappa_0}{2}\left(\left(n_u - g_{u\to d}\left(\frac{l_u}{l_{eq}}\right)\right)(T_S^2 - T_0^2) + g_{u\to d}\left(\frac{l'_d}{l_{eq}}\right)(T_{PM}^2 - T_0^2)\right), \tag{S19}$$

where $l_u$ is the distance the upstream edge modes propagates from S to PM, and $l'_d$ is the distance that the downstream mode propagates from PM to G.

In our devices, $l_u = l_d$, and $l'_u = l'_d$. In addition, the inter-mode interaction is invariant under $n_d \leftrightarrow n_u$ implying $g_{d\to u} = g_{u\to d}$. As a result, the subtraction of the impinging power in the two geometries is independent of the equilibration length and yields the topological thermal Hall conductance,

$$P_d - P_u = (n_d - n_u)\frac{\kappa_0}{2}(T_S^2 - T_0^2). \tag{S20}$$

# 6  POWER CARRIED BY 'DOUBLE-STEP' DISTRIBUTIONS

We derive Eq. 4 in the main text and discuss the conditions we expect it to hold. The most straightforward way to derive the power carried by the fluctuations is to integrate over the entire range of the spectral density (over frequency). When an integer mode carrying current $I$ is partitioned in a quantum point contact (QPC) with probability $t$, the spectral density of the outgoing fluctuations is given by [S11],

$$S(f) = G_0 t(1-t) \left[ (hf + eV) \coth\left(\frac{hf + eV}{2k_b T}\right) \right.$$
$$\left. + (hf - eV) \coth\left(\frac{hf - eV}{2k_b T}\right) - 2hf \coth\left(\frac{hf}{2k_b T}\right) \right]. \tag{S21}$$

The spectrum of the fluctuations is plotted in Fig. S3. Performing the integral over Eq. S21 indeed yields Eq. 3 of the main text

$$P_{AC} = \frac{1}{2G_0} \int_0^\infty df\, S(f) = \frac{G_0}{2} V^2 t(1-t). \tag{S22}$$

A different approach is to calculate directly the energy current carried by the out-of-equilibrium distribution. The energy current (identical to power in 1D) carried by an electronic chiral edge mode with a distribution $f(\epsilon)$ is given by,

$$P = \frac{1}{h} \int_0^\infty d\epsilon\, \epsilon f(\epsilon). \tag{S23}$$

We consider the double step distribution given by the statistical sum of a hot (biased) edge mode with probability $t$ and a cold (unbiased) edge mode with probability $1-t$. The distribution of the cold edge could be described by a Fermi-Dirac distribution with temperature $T$ and chemical potential $\mu_0$ - $f_0(\epsilon) = f_{FD}(\epsilon; \mu_0, T)$. Similarly, the distribution of the hot edge is given by $f_V(\epsilon) = f_{FD}(\epsilon; \mu_0 + eV, T)$, and we assume a linear response $\mu_0 \gg eV, T$. The total excess power is thus

$$P = \frac{1}{h} \int_0^\infty d\epsilon\, \epsilon \big( t f_h(\epsilon) + (1-t) f_c(\epsilon) - f_c(\epsilon) \big) = \frac{1}{2} t G_0 V^2. \tag{S24}$$

And one can derive the power carried by the fluctuations by subtracting the power carried by the DC current $P_{AC} = P - \frac{1}{2G_0}(tI)^2$.

The two calculations above rely on the independent tunneling of electrons across the QPC, forming the double-step distribution, thus generating shot noise with tunneling charge $e$. However, redistribution of the edge modes [S6] and the melting of double-step distributions [S12] has been observed, as well suppressing shot noise. Nonetheless, Eq. 3 in the main text is valid as long as energy and current are conserved. To see it, let us consider energy conservation in the QPC, implying that the incoming power must be equal to the outgoing power (transmitted and reflected),

$$P_{DC}^{in} = P_{DC}^{t} + P_{AC}^{t} + P_{DC}^{r} + P_{AC}^{r}, \tag{S25}$$

where $P_{\text{DC}}^{\text{in}} = \frac{1}{2G_0}I^2$ is the incoming power to the QPC, $P_{\text{DC}}^{\text{t}} = \frac{1}{2G_0}(tI)^2$ is the DC power carried by the transmitted mode, $P_{\text{DC}}^{\text{r}} = \frac{1}{2G_0}\big((1-t)I\big)^2$ is the DC power carried by the reflected mode, and $P_{\text{AC}}^{\text{t}}$ ($P_{\text{AC}}^{\text{r}}$) is the power carried by the fluctuations in the transmitted (reflected) mode. Current conservation in all frequencies implies that the transmitted and reflected noises are anti-correlated (the total outgoing AC current must be zero), and in particular,

$$P_{\text{AC}}^{\text{t}} = P_{\text{AC}}^{\text{r}} = \frac{1}{2}\big(P_{\text{DC}}^{\text{in}} - P_{\text{DC}}^{\text{t}} - P_{\text{DC}}^{\text{r}}\big) = \frac{1}{2G_0}t(1-t)I^2. \tag{S26}$$

The observation of smaller power in the inner edge mode of $\nu = 2$ implies the violation of either energy conservation (due to energy leak to reconstructed neutral modes [S13]), or current conservation (perhaps of the high-frequency fluctuations).

## 7 AMPLIFIER CALIBRATION

When the sample is in equilibrium (no electrical or thermal bias), the spectral density of the voltage noise has two contributions,

$$S_V = A^2(S_{\text{base}} + S_{\text{JN}}), \tag{S27}$$

where $S_{\text{base}}$ is the amplifier noise, $A$ is the gain, and $S_{\text{JN}} = 4k_b TR$ is the equilibrium Johnson Nyquist noise [S14, S15]. By linear fitting $S_V$ with the cryostat's temperature $T$ (at temperatures larger than $30mK$) we extract both $A$ and $S_{\text{base}}$ (Fig. S5). The amplifier's base noise is used to calculate the electron's temperature when the cryostat reaches its lowest temperature (usually at temperatures below $T \sim 20mK$ the temperature of the electronic bath decouples from that of the cryostat duo to the week electron-phonon interaction). Using Eq. S27 we find $T_0 = \left(\frac{S_V}{A^2} - S_{\text{base}}\right)/4k_b R$.

| Device | Chip design | $\nu$ | $T_0$ (mK) | Channel configuration | $\kappa/\kappa_0$ |
|---|---|---|---|---|---|
| Device 1 Density: $1.1 \times 10^{11} cm^{-2}$ Mobility: $4.0 \times 10^6 cm^2 V^{-1} s^{-1}$ | 3 arms Distance $20\mu m$ | 2 | 11 | DS 2 | $1.91 \pm 0.04$ |
| | | | | DS 1 | $0.97 \pm 0.03$ |
| | | | | 0 | $-0.01 \pm 0.03$ |
| | | | | US 2 | $0.01 \pm 0.04$ |
| | | $\frac{2}{3}$ | 17 | DS | $0.41 \pm 0.06$ |
| | | | | US | $0.34 \pm 0.04$ |
| | | | | 0 | $0.01 \pm 0.03$ |
| Device 2 Density: $1.4 \times 10^{11} cm^{-2}$ Mobility: $4.3 \times 10^6 cm^2 V^{-1} s^{-1}$ | 2 arms Distance $20\mu m$ | 2 | 11 | DS 2 | $1.95 \pm 0.03$ |
| | | | | DS 1 | $0.97 \pm 0.03$ |
| | | | | 0 | $0.00 \pm 0.01$ |
| | | 3 | 10 | DS 3 | $2.69 \pm 0.07$ |
| | | | | DS 2 | $1.80 \pm 0.04$ |
| | | | | DS 1 | $0.84 \pm 0.03$ |
| | | $\frac{2}{3}$ | 13 | DS 1 | $0.33 \pm 0.02$ |
| | | | | US 1 | $0.36 \pm 0.03$ |
| | | | | 0 | $0.02 \pm 0.02$ |
| | | | 34 | DS 1 | $0.37 \pm 0.04$ |
| | | | | US 1 | $0.36 \pm 0.04$ |
| | | | | 0 | $0.02 \pm 0.02$ |
| | 3 arms Distance $10\mu m$ | 1 | 30 | DS 1 | $1.07 \pm 0.04$ |
| | | | 71 | DS 1 | $1.14 \pm 0.28$ |
| | | $\frac{2}{3}$ | 13 | DS 1 | $0.42 \pm 0.03$ |
| | | | | US 1 | $0.39 \pm 0.04$ |
| | | | | 0 | $0.03 \pm 0.02$ |
| | | | 30 | DS 1 | $0.41 \pm 0.09$ |
| | | | | US 1 | $0.32 \pm 0.11$ |
| | | | 51 | DS 1 | $0.52 \pm 0.13$ |
| | | | | US 1 | $0.4 \pm 0.15$ |

| Device 3 Density: $3.2 \times 10^{11} cm^{-2}$ Mobility: $1.1 \times 10^7 cm^2 V^{-1} s^{-1}$ | 3 arms Distance $10 \mu m$ | 2 | 16 | DS 2 | $2.09 \pm 0.08$ |
|---|---|---|---|---|---|
| | | | | DS 1 | $1.08 \pm 0.05$ |
| | | | | 0 | $0.12 \pm 0.09$ |
| | | $\frac{4}{3}$ | 12 | DS 2 | $1.98 \pm 0.06$ |
| | | | | DS 1 | $0.97 \pm 0.05$ |
| | | | | 0 | $0.14 \pm 0.03$ |

**Table S1. Reproducibility of the results.** Thermal conductance is measured in all devices and geometries. The channels' configuration corresponding to the channels that conduct the heat from the source to the PM; DS – downstream, US – upstream, 0 – no modes. The number (0,1,…) refers to the number of innermost modes.

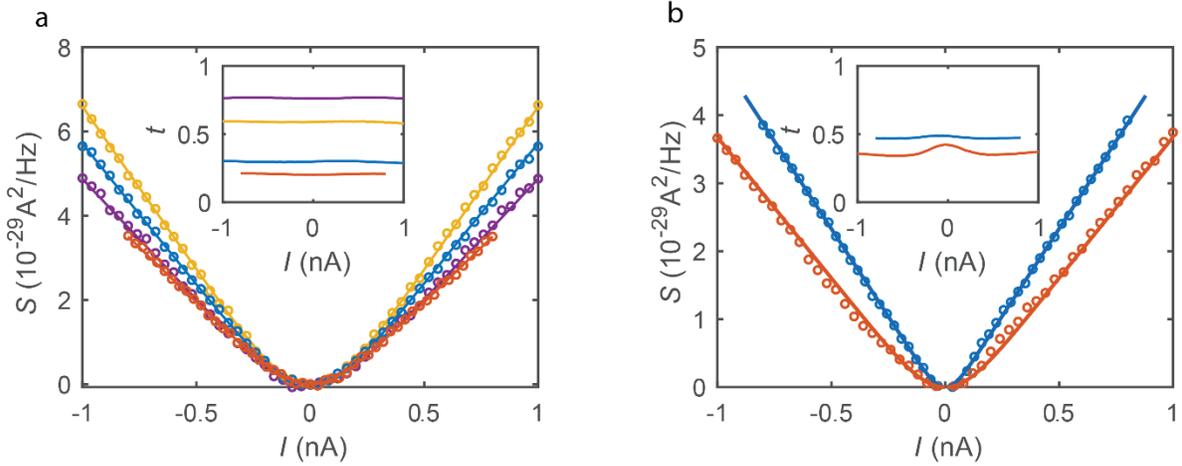

**Figure S1. Shot noise.** Low frequency shot noise as a function of the partitioned current $I$. The different colored markers correspond to different transmissions $t$ of the QPC. The differential transmission is plotted in the inset. The straight lines are fits to the data according to Eq. S1 (**a**) Outer edge mode. We find for $t = 0.76$ (purple) $e^\star/e = 1$ (by calibration) $T = 22 \pm 3 mK$; for $t = 0.59$ (yellow) $e^\star/e = 0.99 \pm 0.01$, $T = 22 \pm 2 mK$; for $t = 0.3$ (blue) $e^\star/e = 1.01 \pm 0.01$, $T = 22 \pm 2 mK$; for $t = 0.21$ (orange) $e^\star/e = 1.01 \pm 0.01$, $T = 19 \pm 2 mK$. (**b**) Inner edge mode. We find for $t = 0.47$ (blue), $e^\star/e = 0.65 \pm 0.01$, $T = 7.0 \pm 0.5$; for $t = 0.38$ (orange), $e^\star/e = 0.55 \pm 0.02$, $T = 17 \pm 2 mK$.

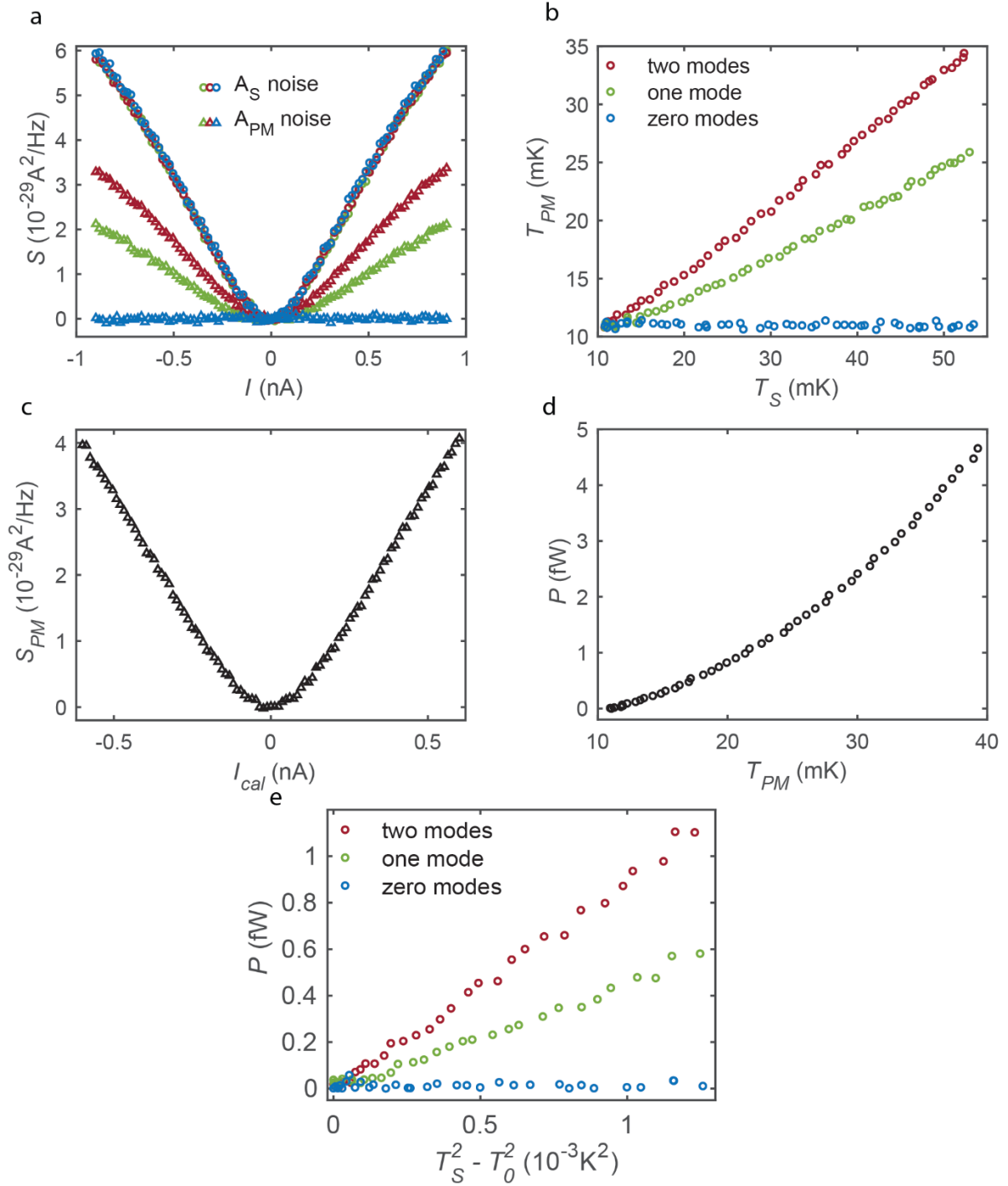

**Figure S2. Measurement protocol for thermal conductance.** (**a**) Measured raw data. Noise picked up at $A_S$ (circles) and $A_{PM}$ (triangles) as a function of the current $I$ sourced from $S_1$ (while current $-I$ is simultaneously sourced from $S_2$). The different marker colors correspond to the number of modes transmitted across GD from S to PM; red – two

modes; green – one mode; blue – zero modes. As expected, the noise at $A_S$ is independent of the number of transmitted modes. (**b**) PM's temperature as a function of the source's temperature for the three different GD configurations. (**c**) Raw data for the calibration process. Noise measured at $A_{PM}$ as a function of the current $I_{cal}$ sourced from $S_1^{cal}$ (while current $-I_{cal}$ is simultaneously sourced from $S_2^{cal}$). (**d**) Calibration of the PM. The impinging power $P = \frac{I_{cal}^2}{G_{2T}}$ as a function of the PM's temperature extracted from the noise. (**e**) Combining the thermal measurement shown in **b** with the calibration measurement shown in **d**, we can plot the impinging power on the PM as a function of the source's temperature (squared).

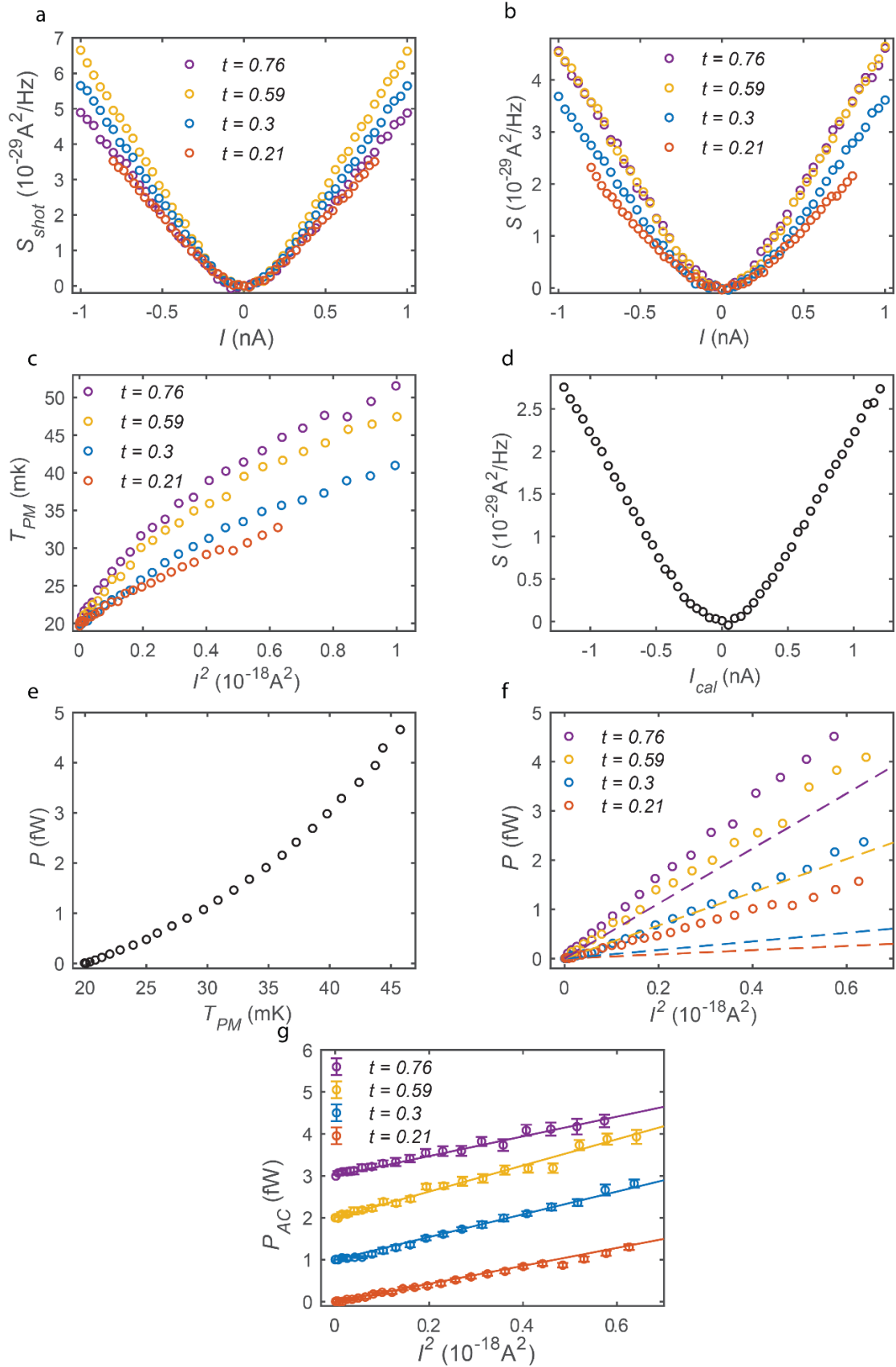

**Figure S3. Measurement protocol for double step distributions.** (**a**) Shot noise measurement. Noise measured at $A_{\text{PM}}$ as a function of the current injected in the outer edge mode. GA is closed. (**b**) Raw data. Noise measured at $A_{\text{PM}}$ as a function of the current injected in the outer edge mode when GA is open. This noise has two contributions – shot noise generated at the QPC and J-N noise generated at the PM. (**c**) By subtracting the shot noise contribution, we can extract the PM's temperature from the noise according to Eq. S9. (**d**) Raw data for the calibration process. Noise measured at $A_{\text{PM}}$ as a function of the current $I_{\text{cal}}$ sourced from $S_{\text{I}}$. Both GD and GA are fully open. (**e**) Calibration of the PM. The dissipated power $P = \frac{1}{8G_0} I_{\text{cal}}^2$ as a function of the PM's temperature extracted from the noise. (**f**) Total power dissipating at the PM as a function of the current injected to the outer edge mode squared, derived by combining the main measurement (plotted in **c**) with the calibration (plotted in **e**). The total power consists of two contributions – the dissipated DC power (plotted as dashed lines) and the power carried by the fluctuations $P_{\text{AC}}$. (**g**) $P_{\text{AC}}$, extracted by subtracting the dissipated DC power from the total power, vs $I^2$. The data is shifted by 1 (fW) for consecutive transmissions for clarity. The straight lines correspond to the theoretically predicted power $P = \frac{1}{2G_0} t(1-t) I^2$ (without any fitting parameters). Here we restored the error bars, marking the statistical scattering of the data.

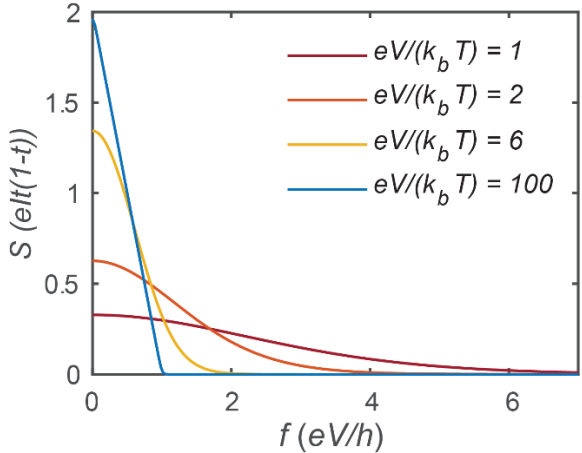

**Figure S4. Noise spectrum.** The spectral density of the noise associated with 'double-step distributions' (in units of $eIt(1-t)$) vs. frequency (in units of $eV/h$) for different temperatures. The power is the area under the graph, and is equal for all the temperatures. The shot noise is the low frequency limit of the curves.

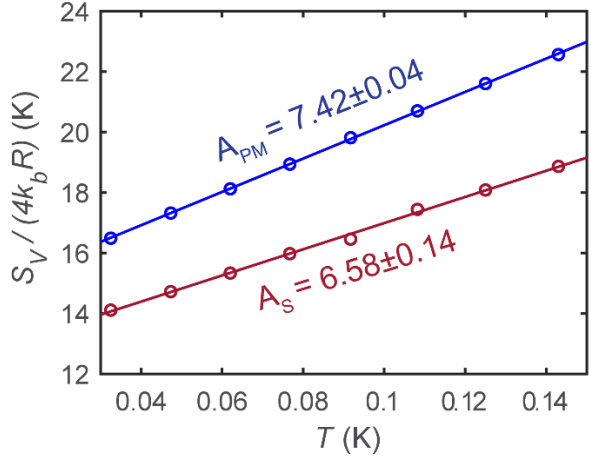

**Figure S5. Amplifier calibration.** Low-frequency spectral density of the voltage fluctuations measured at the two amplifiers (normalized by $4k_\mathrm{b}R$ ) in equilibrium. The background voltage fluctuations linearly increase with the cryostat's temperature. The slope of the linear fit is the gain squared of the amplifier.